\documentclass[12pt,preprint]{aastex}
  
\begin{document}

\title{NGC~2419 -- Another Remnant of Accretion by the Milky Way
\altaffilmark{1}}

\author{Judith G. Cohen\altaffilmark{2},  Evan N. Kirby\altaffilmark{2,3},
Joshua D. Simon\altaffilmark{4} \& Marla Geha\altaffilmark{5} }

\altaffiltext{1}{Based in part on observations obtained at the
W.M. Keck Observatory, which is operated jointly by the California 
Institute of Technology, the University of California, and the
National Aeronautics and Space Administration.}

\altaffiltext{2}{Palomar Observatory, Mail Stop 249-17,
California Institute of Technology, Pasadena, Ca., 91125, 
jlc(enk)@astro.caltech.edu}

\altaffiltext{3}{Hubble Fellow}

\altaffiltext{4}{Carnegie Observatories,
813 Santa Barbara Street, Pasadena, Ca. 91101,
jsimon@obs.carnegiescience.edu}

\altaffiltext{5}{Astronomy Department, Yale University, New Haven,
Ct 06520. marla.geha@yale.edu}

\begin{abstract}

We isolate a sample of 43 upper RGB stars in the extreme outer halo
Galactic globular cluster NGC~2419 from two
Keck/DEIMOS slitmasks.  The probability that there
is more than one contaminating  halo field star in this sample
is extremely low.  Analysis of moderate resolution spectra 
of these cluster members,  as well as of our Keck/HIRES
high resolution spectra of
a subsample of them, demonstrates that there is a small
but real spread in Ca abundance of $\sim$0.2~dex within this massive metal-poor
globular cluster.  This provides additional support to earlier
suggestions that NGC~2419 is the remnant of a dwarf galaxy
accreted long ago by the Milky Way. 

\end{abstract}

\keywords{Galaxy: globular clusters: individual (NGC~2419), Galaxy: formation, 
Galaxy: halo}

\section{Introduction \label{section_intro} }

NGC~2419 is a globular cluster (GC) in the outer halo of the Galaxy
at a distance of 84~kpc\footnote{This and other parameters for
NGC~2419 are from the current version of the on-line GC 
database of \cite{harris96}, based on the photometry of \cite{harris97}.
The first photometric study of this GC was by \cite{racine76}.}. 
It has the highest luminosity ($M_V \sim -9.6$~mag) of
any GC with galactocentric radius $R > 20$~kpc,
higher than all other GCs with $R > 15$~kpc with the exception of M54, which is
believed to be associated with the  nucleus of the Sgr dSph galaxy
now being accreted by the Milky Way \citep{ibata95,sarajedini95}.
NGC~2419 also has an unusually large  half-light radius
($r_h \sim 19$~pc) and core radius
for such a massive Galactic GC, and a  relaxation time which exceeds
the Hubble time, also unusual for a GC.  Although there is no dynamical
evidence for dark matter in NGC~2419 (or any other GC)\footnote{Although
\cite{cote02} deduced an apparently high
$M/L$ for Pal~13,
a low luminosity distant halo GC, from their
measurement of  $\sigma_v \sim 2.2\pm0.4$~km~s$^{-1}$,
\cite{siegel01}
suggest that Pal~13 is ``in its final throes of destruction''
while \cite{blecha04} demonstrate that $\sigma_v$ as determined
in previous work has been inflated by binaries; 
with their new measurement there is
no need for dark matter.}, 
if dark matter was initially present in this object, 
it should not have been depleted from even the
central region of this GC.  Furthermore, given the position
of NGC~2419 in the outer halo, it
should not have been affected
by any tidal interaction with the Milky Way, 
assuming its orbit is not extremely eccentric.
\cite{sigmav09} determined
a velocity dispersion for this cluster based on 40 stars; they found $M/L = 
2.05\pm0.50~M_{\odot}/L_{\odot}$, a normal value for an old
stellar system,  with no evidence for the presence of
dark matter.
\cite{harris97} obtained a deep CMD of NGC~2419 with HST/WFPC2 and 
demonstrated  that there is no detectable difference
in age between it and M92, a well studied inner halo GC
of comparable metallicity\footnote{A subsequent analysis of all existing HST
data was carried out by \cite{sandquist08} focusing on the characteristics 
of the horizontal branch.}.

Every other luminous GC in the outer halo is considerably more compact.
The current version of the on-line database of \cite{harris96} lists
only three Galactic GCs that lie anomalously above the bulk
of the Galactic GCs 
in the plane $r_h$ vs $M_V$, one of which is
NGC~2419. The  other two are M54 and $\omega$ Cen, 
the most luminous Galactic GC,
and one with a large internal spread in metallicity.  
The M31 cluster G1 also lies  in this anomalous range
\citep{ma07}.  
The latter three of these four GCs,
all of which are all extremely luminous, are believed to be the stripped 
cores of dwarf galaxies accreted by our galaxy or, for G1, by M31.

NGC~2419 is quite metal poor. 
\cite{shetrone01} observed one star in this GC with HIRES/Keck; they
confirm the low
metallicity derived from moderate-resolution spectroscopy,
[Fe/H] $\sim -2.1$~dex,  of \cite{suntzeff88}.  All other
existing spectra of stars in NGC~2419 prior to our work 
are of low S/N, suitable
only for radial velocity determination.

Following the early work of \cite{searle78} and of
\cite{zinn96}, \cite{mackay04} and others have discussed the possibility
that some of the outer halo Milky Way GCs have been accreted,
while \cite{mackey10} argue similarly for the outer halo M31 GCs.
\cite{mackey05}
\citep[see also ][]{vdb04}
suggest that NGC~2419 is such a case, and that it 
is the nucleus of a dwarf galaxy that has been accreted by the Milky Way.
\cite{pm2419}
have measured proper motions and determined a preliminary
orbit for the Virgo stellar stream; they claim that NGC~2419 is
the nucleus of a disrupted stellar system of which the
Virgo stellar stream is a part.

\begin{figure}
\epsscale{0.9}
\plotone{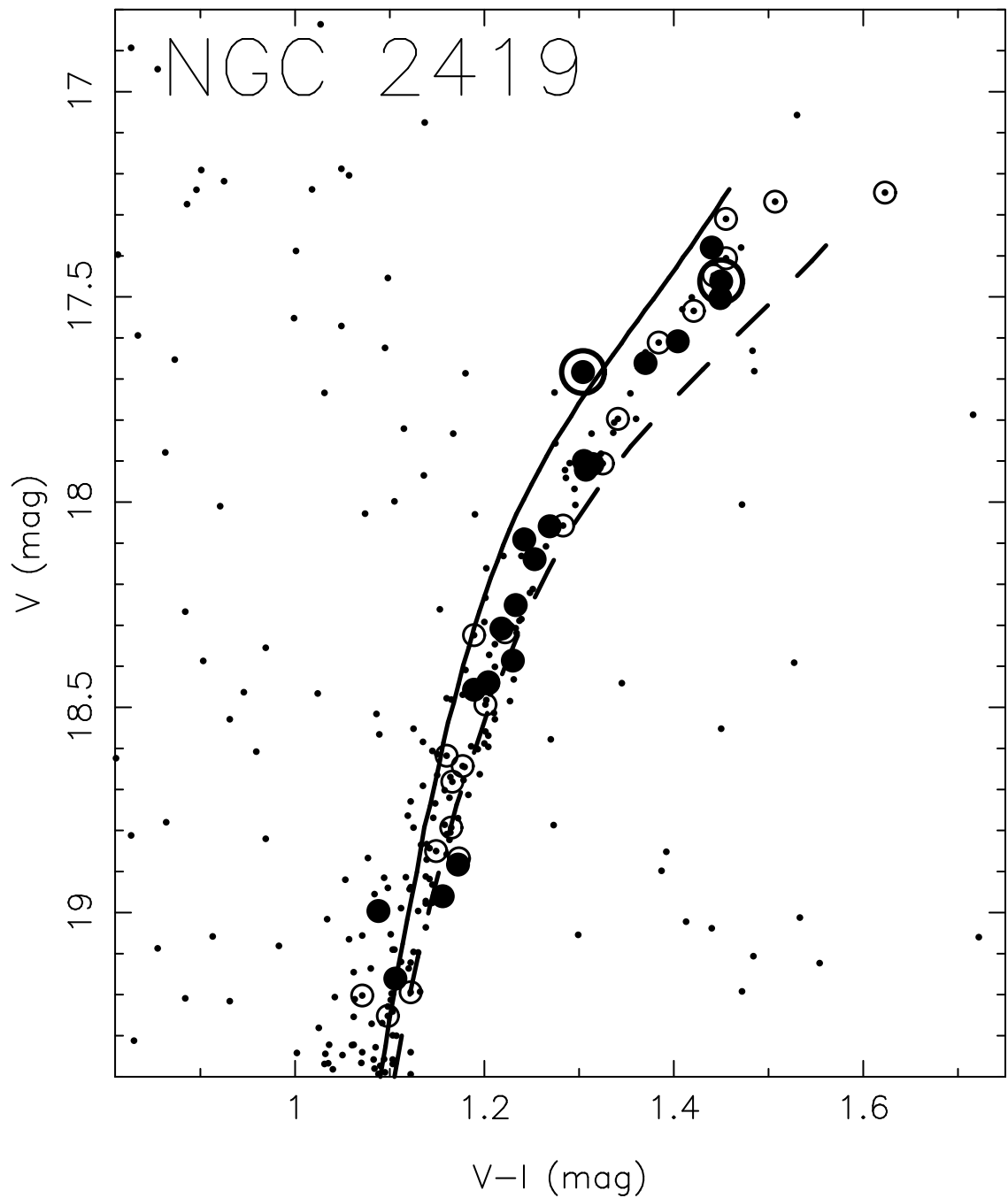}
\caption[]{The V, V--I CMD for the database of \cite{stetson05} in the
field of NGC~2419.  Those stars in the DEIMOS slitmasks judged
to be members of the GC are indicated by open circles (low CaT)
or filled circles (high CaT).  The two stars whose membership status
is still uncertain
are circled.  Isochrones from \cite{yi03} with
[$\alpha$/Fe] 0.3~dex, age 12~Gyr, and [Fe/H] $-1.90$ (dashed line) and
$-2.2$~dex (solid line) (interpolated using the code they supply) are shown.  
A 0.04~mag offset in $V-I$ was applied
to both isochrones to improve the fit to the observations.
\label{figure_cmd}}
\end{figure}

There is no evidence from the CMD of multiple stellar populations in NGC~2419,
as is illustrated in Fig.~\ref{figure_cmd}.
Because of its low metallicity, the isochrones along the RGB
as a function of [Fe/H] are very close together in color, and
thus it will be difficult to detect a
star-to-star range in [Fe/H] from a measured width in color of the
upper giant branch in this cluster's CMD. 
The observed
range in $B-V$ at a fixed $V$ on the upper RGB 
in the ground based photometric database of 
\cite{stetson05}, on which we rely
extensively,  in the field of NGC~2419 is $\sim$0.05~mag, somewhat larger 
than the expected observational
uncertainty.  If this in total  is ascribed to 
metallicity variations,
it corresponds to a spread in  [Fe/H] of $\sim$0.2 dex.
The constraint on potential internal variations of [Fe/H]
within NGC~2419 set by the width of the upper RGB from both
ground and  HST observations 
is comparable
as V--I is less sensitive to metallicity variations than is B--V.

The existence of multiple populations in GCs has been known
to spectroscopists for decades, see, e.g. \cite{m15cn}
and more recently \cite{carretta09}.
As reviewed by \cite{piotto09},
the  exquisite photometry from the HST/ACS Survey
of Galactic Globular Clusters \citep{sarajedini07}
has enabled the detection of subtle features in GC CMDs 
that were previously lost in the noise.  These  include 
multiple main sequences, a double
subgiant branch in NGC~1851 \citep{milone08}, and other such 
phenomena suggesting the presence of multiple populations.

They are explained by small differences in age, He content
(which, although difficult to detect directly, must occur in association 
with the ubiquitous C/N and O/Na  anti-correlations seen in all GCs), 
and/or heavier metal content.  When invoking the latter, one must be careful
to distinguish between variations among those light elements
that can be
produced by ejecta from intermediate or low mass AGB stars,
versus variations in the heavier elements such as Ca or Fe, which
are only produced in SN. 

The issue of a second generation of
SNII contributions is of particular interest, since it is hard
to understand how a  GC with little or no dark matter
and low binding energy
could retain energetic SN ejecta, unless the GC we observe
today were a remnant of an initially much more massive stellar system.
From both spectroscopy and photometry $\omega$~Cen has been known 
for more than 30 years to have a wide
intrinsic range in [Ca/H], [Fe/H] etc.
extending over a range of $\sim$1.3~dex with
multiple peaks \citep{norris96}.
The GC M54 also shows this, but it is believed to be part of the central
region of the Sgr dSph galaxy, currently being accreted by the Milky Way.
Very recently M22, under suspicion for many years, was confirmed
to have such star-to-star variations in Ca/H
and/or Fe/H by three independent groups,
\cite{costa09}, \cite{marino09}, and \cite{lee_ca3933}, who 
each find a range in [Fe/H] and/or [Ca/H] of 
$\sim$0.2~dex. 
In the present paper we explore whether
NGC~2419 could be another example of a massive GC which has 
multiple populations with a range
in heavy element abundances. 

\section{Observations \label{section_obs} }

We have obtained two multi-object slitmasks of NGC~2419
with the DEIMOS spectrograph \citep{faber_deimos} at the Keck~II telescope
covering the range $\sim 6500$ to $\sim 9200$~\AA, thus including
the near infrared Ca triplet.
The first of these was used by J.~Simon and M.~Geha to
evaluate their radial velocity accuracy for studies
of ultra-faint dwarf spheroidal satellites of the Galaxy \citep{simon07}.
Three 600 sec exposures of this slitmask were taken on  Feb. 2, 2006
and three 660 sec exposures followed on Feb 4, 2006. The
spectra from each night were summed.
Spectra of 65 stars were extracted of which about 35
have adequate S/N and appear
to be metal-poor red giants which might be members of NGC~2419.

In early 2009, J.~Cohen  collected from various people
(including J.~Simon and E.~Kirby)
spectra from DEIMOS slitmasks of GC fields that include the
8500~\AA\ region.  From an initial exploration of such data
for 6 GCs,
NGC~2419 was quickly isolated as a potentially interesting case,
as even a preliminary analysis suggested there might be a small
spread in Ca abundance based on the strength
of the near-IR Ca triplet.  Cohen and Kirby 
then obtained a second DEIMOS slitmask covering NGC~2419, as
well as  similar data for several other GCs, see \cite{kirby10}.
The second NGC~2419 slitmask was used at the Keck~II telescope in
Oct. 2009 on two consecutive nights for a total of exposure time of 4800~sec
and yielded 96 spectra.
The position angle of the 2009 slitmask was 178$^{\circ}$
different from that of the 2006 slitmask, and the offset between
the centers was 42~arcsec.  The slitmasks were placed as far
as possible from HD~60771, with $V \sim 7.2$~mag, located
about 240~arcsec East of the cluster center.  Any scattered light
from this bright A5 star would appear as a
contribution to the diffuse night sky, and would be removed via
subtraction of the night sky, but inspection of the raw DEIMOS frames
showed there is no noticeable background from the star.

Both of these slitmasks were reduced with the DEEP 
pipeline\footnote{See
http://astro.berkeley.edu/$\sim$cooper/deep/spec2d.}.
The resulting spectra were then processed
through a Figaro \citep{figaro} script which determined
a continuum level as a function of wavelength, then measured
pseudo-equivalent widths through bandpasses defined for
the two stronger lines in the near-IR Ca triplet at
8542 and 8662~\AA.  These were defined as 
$ [1.0 - (\sum c_i/N_{pix})] \Delta\lambda$
where $c_i$ is the continuum-normalized DN/pixel,
$\Delta\lambda$ is the bandwidth used for the feature,
and $N_{pix}$ is the number of pixels within this bandpass.  
The weakest triplet line at 8498~\AA\
was ignored due to the low metallicity of the GC of most
interest, NGC~2419.  DEIMOS slitmasks of 8 other Galactic GCs were
analyzed in the same way with scripts modified only to take
into account the differing radial velocities of each GC
and extending the index bandpass widths to accommodate
the increasing strength of the Ca triplet lines for more metal-rich
GCs.

The two slitmasks for NGC~2419 combined 
yielded 
moderate resolution spectra of more than 130 stars.
Those which upon visual inspection were not
metal-poor giants, including stars appearing too metal-rich, M dwarfs,
or stars with strong Paschen lines, were rejected.  In addition,
those which were discrepant from the cluster mean
$v_r$ by more than 15~km s$^{-1}$ as determined from cross 
correlations  were rejected.
A further requirement imposed was that
a candidate member be within 0.1~mag of the $V,V-I$ locus of the 
RGB of NGC~2419 as defined from the photometric database of
\cite{stetson05}.
This left a sample of 45
apparent RGB members of NGC~2419.
Fig.~\ref{figure_4gc_cat} shows
the sum of the pseudo-equivalent widths for the two
stronger Ca triplet lines, which we denote CaT, 
measured in this way from the two slitmasks of NGC~2419,
with two possible non-members omitted (43 RGB stars in total), together
with large samples from the well-studied Galactic GCs M15, M22, and M5.
Note the apparent spread in CaT at a fixed $V$~mag of the NGC~2419 sample, 
also shown by the M22 sample, but not by the M15 or M5 giants.

The CaT measurements for the 43 NGC~2419 members isolated from
our Deimos slitmasks have been converted into [Ca/H] values,
and their histogram is shown in Fig.~\ref{figure_feh_hist}.
In this exercise, the calibrating GCs we used are
M15, M53, and M13, with [Fe/H] assumed to be $-2.26$,
$-1.99$, and $-$1.54~dex respectively, adopted from the on-line
catalog of \cite{harris96}.  Furthermore, 
based on the cumulative
result of the many studies of GCs reviewed
and summarized by \cite{gratton_araa},
we assume [Ca/Fe] 
is constant at +0.25~dex for all of the red giants
in the calibrating GCs. The 
sensitivity of Ca triplet line strength
to [Ca/H] is strong in this low metallicity regime 
([Fe/H] $\lesssim -2$~dex);
CaT changes by 0.1~\AA\ for a 0.02~dex change in [Ca/H]. 
Fig.~\ref{figure_feh_hist}
shows that [Ca/H](CaT) in NGC~2419 is stongly peaked
at about $-1.9$~dex, with a tail of members extending
to somewhat higher metallicity.  The exact peak metallicity
for NGC~2419
depends on the [Fe/H] values assigned to the calibrating
GCs and on the constant [Ca/Fe] adopted. The latest detailed abundance analysis for M15 
luminous giants by \cite{carretta09} finds [Fe/H] $-2.32$~dex,
0.06~dex below the value adopted here; their somewhat lower [Fe/H]
for this GC is supported by
as yet unpublished work by J.~Cohen and J.~Melendez.
M53 is the least well studied at high dispersion of the calibrators 
and thus might have a value for [Fe/H] in the \cite{harris96}
database which is incorrect by up to $\pm$0.15~dex


Even from just the 2006 slitmask, it was clear that NGC~2419
showed signs of having a spread in Ca/H.    A campaign was
therefore launched to obtain high resolution spectra of
a number of RGB stars in this GC, with 8 stars observed 
with HIRES \citep{vogt94} at the Keck~I telescope, of which  5
have reasonable S/N, while the other three, observed
with HIRES-B  during a run for another program, are somewhat lower in S/N.
In addition, several somewhat fainter stars suspected
to have unusually strong Ca triplet lines from the moderate
resolution DEIMOS spectra were observed
with HIRES in Feb. 2010 with typical exposures of 1800 sec
to get accurate radial velocities.

\begin{figure}
\epsscale{1.0}
\plotone{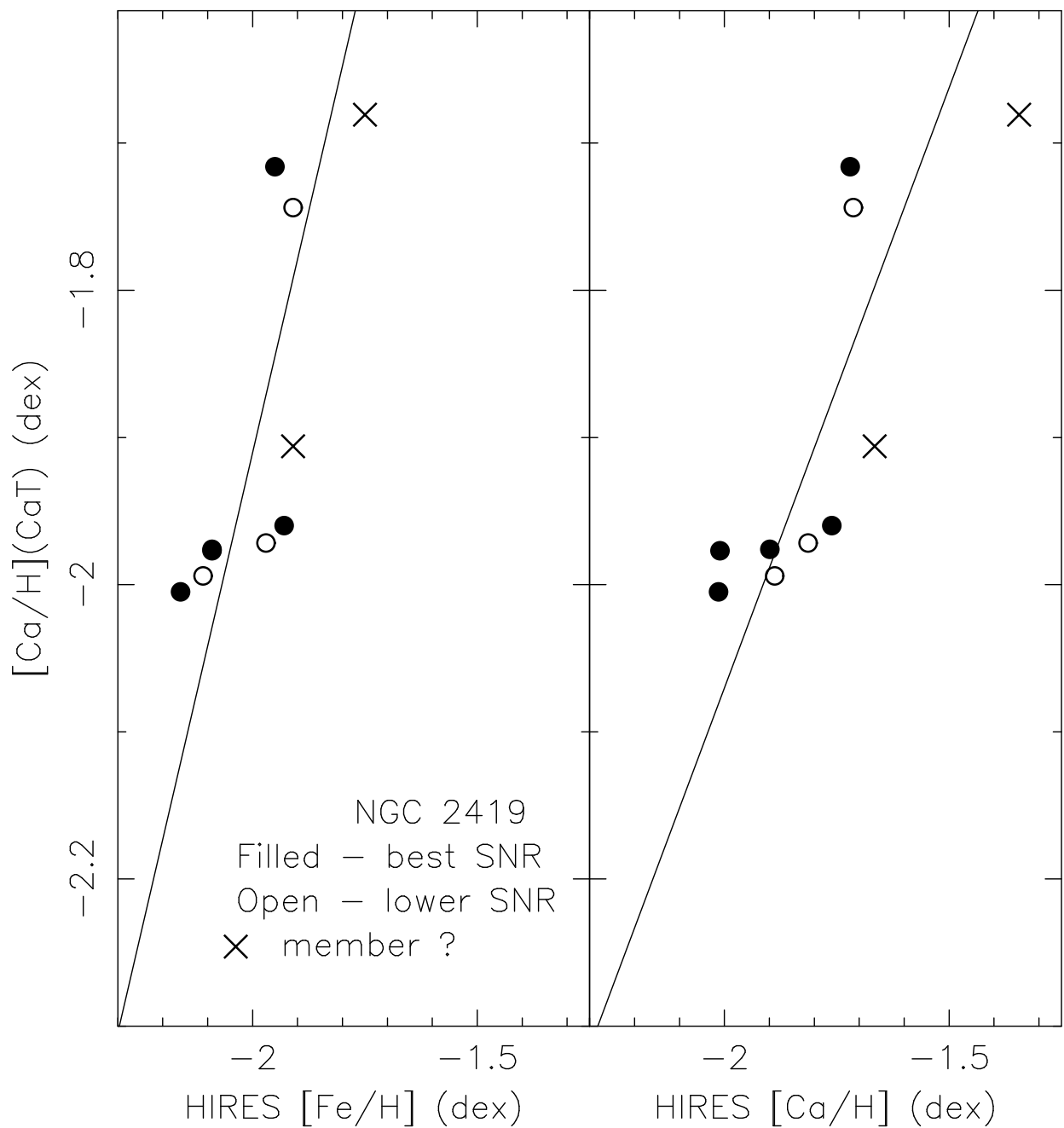}
\caption[]{[Ca/H] derived from the sum of the pseudo-equivalent widths for the two stronger lines
of the near-infrared Ca triplet are shown as a function of
the [Fe/H] (left panel) or [Ca/H] (right panel) for the NGC~2419
stars with HIRES spectra.  Giants with higher S/N spectra
are indicated as filled circles; open circles denote
the members with lower S/N spectra, two of which,
marked by crosses, still have concerns
regarding membership (see \S\ref{section_member}). 
\label{figure_cah_hires_cat}}
\end{figure}

The HIRES spectra were analyzed in a manner similar to that
described in \cite{8_new_emp} with one major difference
intended to improve the relative accuracy of the determination
of $T_{eff}$ for each star;
we explicitly assumed that the NGC~2419 RGB stars 
all lie on a single isochrone.
Full details of the abundance analysis will be given
Cohen et al (2010, in preparation). 
Fig.~\ref{figure_cah_hires_cat} compares the resulting derived
[Fe/H] and [Ca/H] for each of the ten candidate members of
NGC~2419 with detailed
abundance analyses to their [Ca/H] as measured from the DEIMOS slitmasks.
It is clear that the [Ca/H]CaT (from DEIMOS spectra) are correlated
with the [Ca/H] from the HIRES spectra, and that both
data sets suggest that a small range in Ca abundance exists
among the upper RGB stars in NGC~2419.  [Ca/Fe]
for  the 8 members with HIRES spectra is $+0.17$~dex
with $\sigma = 0.05$~dex; the total range in [Fe/H] and in [Ca/H]
is 0.25 and 0.31~dex respectively, which ranges are identical
to within the errors.

\section{Membership \label{section_member}}

Establishing the membership of
candidates in NGC~2419 is complicated as the cluster mean 
$v_r$ is  $\sim -20.3$~km s$^{-1}$,
which is not very far from the mean of the halo field
at about 0~km s$^{-1}$.  
Furthermore, the metallicity
of NGC~2419 is close to that of the mean of field
stars with kinematics suggesting their orbits reach the outer halo
of [Fe/H] $\sim -2.0$~dex \citep{carollo07},
so confusion with any halo field
giants whose distances are comparable to that of the GC may occur.
Moderate resolution spectra can rule out
only those stars which have $v_r$ or CaT considerably different from
the cluster mean.  The radial velocity uncertainty from
the  DEIMOS spectra for NGC~2419 stars in the relevant magnitude range
was determined by comparing the $v_r$ measured from each of the
two slitmasks, taken three years apart, for the 19 stars in common.
The result was a difference in the mean of 0.01~pixels with
$\sigma$ less than
0.2~pixels on the detector; this corresponds to  
$\sigma \sim 2.3$~km s$^{-1}$.

With regard to the initial list of 10 discrepant stars
with CaT somewhat high for their $V$ from Fig.~\ref{figure_4gc_cat},
three
were already confirmed as members from the high resolution spectra
of \cite{olszewski93} or of \cite{sigmav09}.  All but one
of the remaining 7 were observed with
short exposures with HIRES in Feb. 2010.  Of these,
the status of NGC~2419~S951 remains unclear. Even though it lies 
on the cluster locus in the CMD, and has a spectrum which appears
reasonable for a member, its $v_r$ is 15~km s$^{-1}$ lower than the
cluster mean $v_r$; the velocity dispersion in this GC is
about 4~km s$^{-1}$ \citep{sigmav09}.  This star could
be a binary in this GC, but the two spectra taken three years apart
have the same $v_r$ to within the uncertainties, or it could
simply be far in the tail of the $v_r$ distribution. 
NGC~2419~S1673 has $v_r$ consistent
with membership in this GC, but
it lies 0.06~mag bluer than the RGB locus for this GC.
It also has slightly higher [Ca/H] and [Fe/H] 
derived from its HIRES spectrum than the 8 stars
we judge members of this GC, as is shown in 
Fig.~\ref{figure_cah_hires_cat}.
We suspect that it may be an AGB member, as it, together with
the most luminous NGC~2419 RGB stars, shows strong emission
in both the red and blue wings of H$\alpha$.  Ignoring these
two stars leaves a sample of 43 RGB stars expected to be members
of NGC~2419.

\begin{figure}
\epsscale{1.0}
\plotone{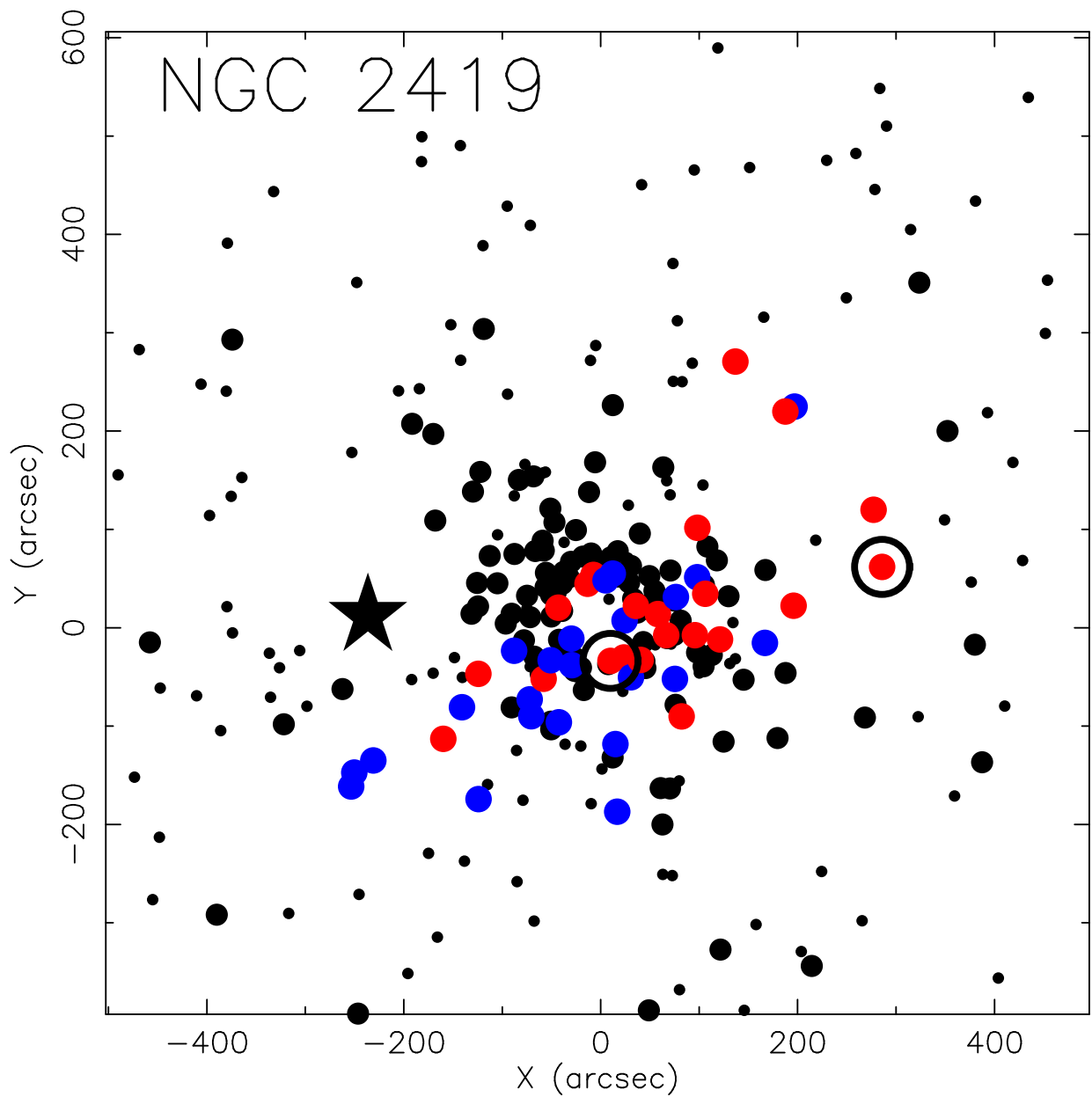}
\caption[]{The positions on the sky of the stars from the photometric
database of \cite{stetson05} are indicated by
small black dots.  Those  
 with $17.2 < V < 19.2$~mag and within 0.2~mag
of the NGC 2419 locus in the $V, V-I$ CMD are shown by
larger filled circles.  Objects in the DEIMOS slitmasks
with low CaT (blue circles) and those with higher CaT(red circles)
for their V~mag are indicated.  The two circled red points
are apparent members whose membership is still uncertain
(see \S\ref{section_member}); NGC~2419 S951 is the one close to the center
of the cluster, while S1673 is far from the center.
The position of HD~60771 is indicated by a large star.
\label{figure_cah_xy}}
\end{figure}

Fig.~\ref{figure_cah_xy} shows the region of NGC~2419 on
the sky with the positions of the stars in our sample indicated, as well
as those of the entire Stetson photometric database in this field
and that of HD~60771.
This figure suggests that the field star
contamination of our NGC~2419 sample of upper RGB stars
should be small.
To quantify this we first
consider the observed star counts in 
a region on the sky within the Stetson
photometric database for NGC~2419 but far from the cluster center.
Within an area of 69 arcmin$^2$, there are no stars
which lie within 0.1~mag of the NGC~2419 locus in the
$V,~V-I$ CMD with $17.2 < V < 19.2$~mag.  This
yields a maximum background rate of
0.01 field stars/arc min$^2$. 
Since there is considerable overlap in the area of
the two DEIMOS slitmasks, we set the total area we used
to 100 arc min$^2$ to predict a maximum of 1 field star.
We then require that $v_r$ must be
that of the mean for this GC to within  10 km~s$^{-1}$;
NGC~2419 S951, whose membership status is still uncertain
as discussed above,
is the only star between 10 and 15 km~s$^{-1}$ off the cluster mean
that survived the initial series of checks.
We adopt the values for the ``outer halo'' of \cite{carollo10} of
$<v_R> = -8.6 \pm 6.1$ km~s$^{-1}$, 
$\sigma(v_R) = 159 \pm 4$ km~s$^{-1}$,
noting that these apply to a region at much smaller galactocentric
radius than the one under consideration here.
Then our criterion for $v_r$ (assumed equal to $v_R$, where $R$ is the galactocentric radius)
rejects 95\% of the field halo stars.  This cuts the expected
number of field stars in our sample to less than 0.005 field stars/arc min$^2$.
Fig.~\ref{figure_cah_xy} also suggests there may be a spatial assymetry
in the distribution of the high CaT vs the low CaT RGB stars
in NGC~2419.

The expected field star contamination was also evaluated
using the Besan{\c{c}}on model of the Milky Way
\citep{besancon}.  With the rough 
color selection used to design the 2006 slitmask (which included a 
somewhat broader box than just within 0.1~mag of the cluster RGB),  
the surface density of halo giants meeting the velocity cut of $-35$ km~s$^{-1} < 
v_r < -5$~km~s$^{-1}$ is predicted to be
to be $\sim$0.001 stars/arc min$^2$.  Given the
radius for the most distant RGB candidate member of NGC~2419 of 
500'', the expected number of halo giants in the observed area is $\sim$0.2.  
One  could then apply an additional metallicity cut to select only stars within 
some range around the cluster [Fe/H] value, but we conservatively 
assume that all of the velocity and color-selected stars will also lie in 
the right metallicity range. 
One must recognize that the Besan{\c{c}}on 
model for the outer halo is probably
not well calibrated at such large galactocentric radii. 
Furthermore, it doesn't have substructure, which is almost certainly 
dominant at large radii.

Since both of these tests predict a number of field star
interlopers in our carefully selected sample of  less than one,
we conclude that the large population in the DEIMOS
sample in the field of NGC~2419
of RGB stars with somewhat higher CaT must  almost entirely 
be members of this GC.

\section{Accuracy of CaT}

We next demonstrate that the range in CaT at a fixed $V$ near the
tip of the RGB shown in Fig.~\ref{figure_cah_hires_cat} is
real.  NGC~2419 is the most distant GC shown in this figure.
But the M5 sample extends more than 5~mag below the
RGB tip of this GC, comparable to the difference in distance modulus
between M5 and NGC~2419.  With only a minimal effort to 
remove non-members
the M5 sequence is much tighter even at the faintest levels probed.
The bulk of the M22 sample is also several magnitudes below
the RGB tip, but does not reach as faint as the NGC~2419 giants.

The DEIMOS spectra have a resolution $\lambda/\Delta\lambda \sim 7000$
with 0.32~\AA/pixel near 8500~\AA.
Our criterion for a suitable S/N DEIMOS spectrum for a candidate member
of NGC~2419 is a total of 3000 detected electrons/pixel
in the continuum at 8600~\AA.  Thus, ignoring issues of sky subtraction
of night sky emission lines, which fortunately are few and fairly
weak in the region of the near-infrared Ca triplet, the S/N
per each 1.3~\AA\ spectral resolution element
in the continuum at 8600~\AA\ will exceed 110 as each of the
two slitmasks has a total exposure time of more than one hour.

We used a width of
25~\AA\ for the 8542~\AA\ line index and of 20~\AA\ for
the weaker 8662~\AA\ line index  for candidate members of NGC~2419,
for a total bandpass for the CaT measurement of 
45~\AA, which is $\sim$34 spectral resolution
elements for the two Ca~II lines.   A total bandpass of 350~\AA\
was used in the continuum fitting, corresponding to 290 spectral
resolution elements.

The range in CaT at a fixed $V$ near
the RGB tip in NGC~2419 is about 1~\AA.  Thus a 2.2\% change in the
continuum level across the entire CaT region could produce the
observed width.  However, our spectra were analyzed with scripts
in a fully automatic manner.  No manual intervention occurred.
Given the S/N requirement for the DEIMOS spectra described above,
it is difficult to understand how an error in the
continuum location systematically either high or low at the level 
required could have occurred.  Beyond the issue of scattered light
from the nearby bright star HD~60771, discussed  and dismissed earlier in
\S\ref{section_obs}, we know of no other issues that might
introduce an apparent spread for just NGC~2419 among our sample
of Galactic GCs with Deimos slitmasks. 

But we have more quantitative demonstrations of the accuracy
of our CaT measurements for NGC~2419.  There are two slitmasks
in this GC taken three years apart, with considerable overlap between them.
Also the two sequences of exposures taken two nights apart
of the 2006 slitmask 
were each reduced independently. The CaT values were measured
from each, then averaged.  The  mean ($\sigma$) of the differences 
between CaT from the first exposure
and that from the second 
is 0.08~\AA~(0.13~{\AA})
for the 27 stars from the 2006 DEIMOS slitmask
which were accepted as members of
NGC~2419 with $V < 19.3$~mag.  Considering the 23  stars brighter
than $V = 18.8$~mag, this is reduced to 0.05~\AA\
with $\sigma = 0.11$~\AA. 

There are 16 stars in common
between the two slitmasks. The
mean of the differences 
between CaT from each of the two slitmasks is 
0.03~\AA\ with $\sigma = 0.13$~\AA. 
This again demonstrates that the spread seen in 
Fig.~\ref{figure_4gc_cat} for the NGC~2419 red giants is 
real. The adopted CaT values 
are averages of those
from the 2006 and the 2009 slitmask, for those stars included
in both slitmasks, and are given in Table~\ref{table_sample}.
 
\subsection{Comparison With Spectral Synthesis of the Deimos Spectra}

\begin{figure}
\epsscale{1.0}
\plotone{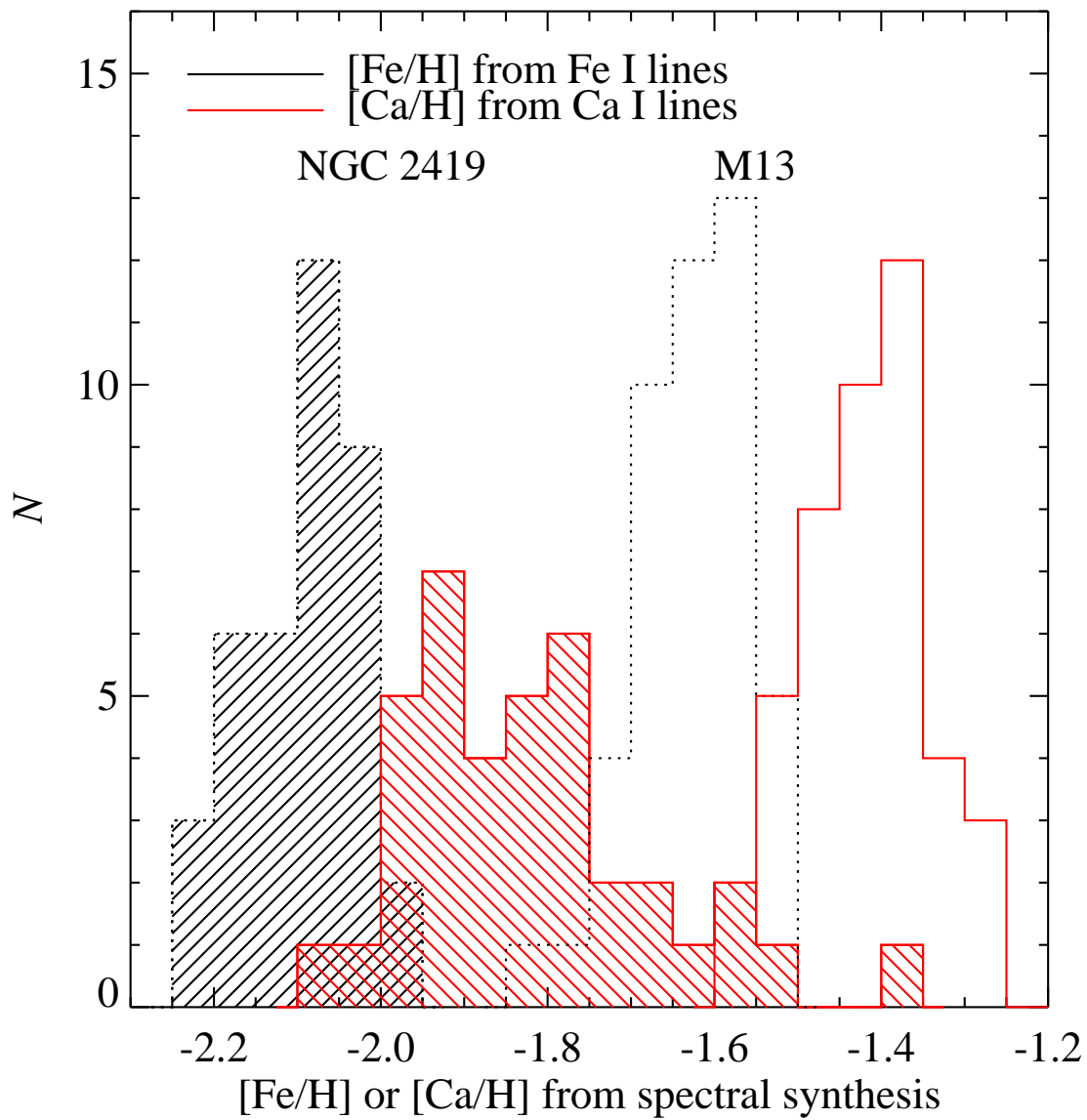}
\caption[]{Spectral synthesis-based [Fe/H] (black) and [Ca/H] (red)
distributions for stars in the globular clusters NGC 2419 (shaded) and M13
(open).  Only stars 
from the list of probable members of this GC isolated here 
(see Table~\ref{table_sample})
with estimated uncertainties of less than 0.2~dex in
both quantities are included.  
\label{figure_kirby_hist}}
\end{figure}

\begin{figure}
\epsscale{1.0}
\plotone{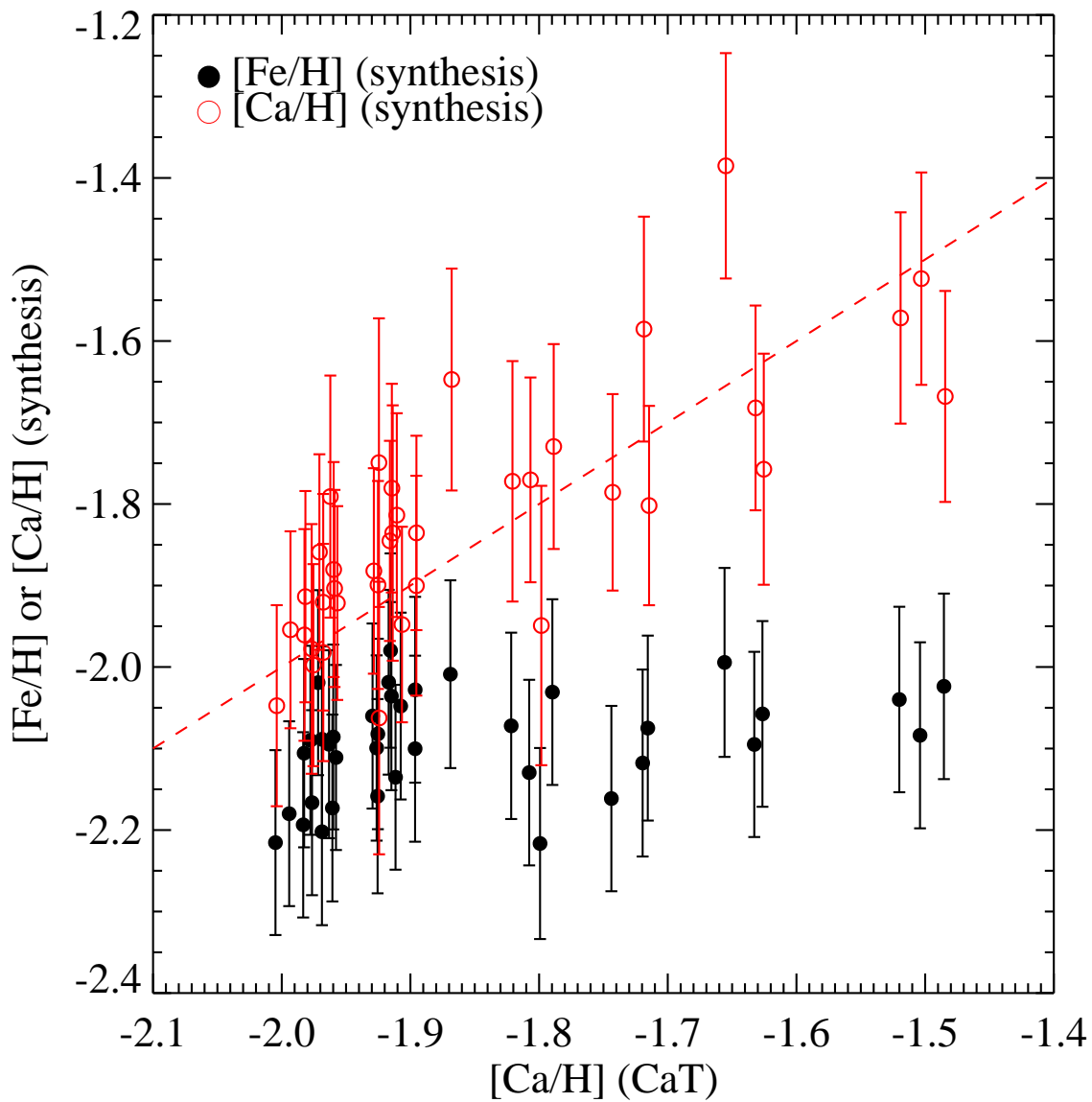}
\caption[]{Spectral synthesis-based [Fe/H] (black) and [Ca/H] (red)
for stars in NGC~2419 
are shown as a function of [Ca/H](CaT). Only stars 
from the list of probable members of NGC~2419 isolated here 
(see Table~\ref{table_sample})
with estimated uncertainties of less than 0.2~dex in
both quantities are included. The dashed line indicates equality between
[Ca/H](CaT) derived from the Ca triplet lines
and that from spectral synthesis excluding the triplet lines.
\label{figure_kirby_starcomp}}
\end{figure}

\cite{kirby08} and \cite{kirby10} have measured [Fe/H] and [Ca/Fe] abundances for
stars in Galactic GCs  and dwarf spheroidal galaxies.  NGC~2419 was
among their sample.  In fact, they used the same DEIMOS data from which we
measure Ca triplet metallicities.  Because their technique is based on
spectral synthesis and because the Ca triplet lines are excluded from their
analysis, their measurements are independent of our Ca triplet-based
results.  Figure~\ref{figure_kirby_hist} shows the 
spectral synthesis-based [Fe/H] and [Ca/H] distribution functions,
where [Ca/H] is the sum of [Fe/H] and [Ca/Fe]  from
the  set of 43 stars isolated as probable members of NGC~2419
here (see Table~\ref{table_sample})
and with [Fe/H] and [Ca/Fe] errors less than 0.2~dex; 38 of the
43 meet all these criteria. 
The spectral
synthesis-based [Ca/H] distribution 
it is asymmetric in the same sense as is that derived from
CaT: a peak
followed by a tail toward higher Ca abundances, as can been seen
by comparing Fig.~\ref{figure_kirby_hist} from the spectral syntheses
vs. Fig.~\ref{figure_feh_hist} from the Ca triplet analysis. 
A comparison for each
individual star in NGC~2419 from our sample that meets these criteria
is shown in Fig.~\ref{figure_kirby_starcomp}.
The dashed line in this figure indicates equality between
the Ca abundances [Ca/H] derived from the Ca triplet lines
and that from spectral synthesis and strongly supports
the reality of the small
range in Ca abundance we claim to exist in NGC~2419.

Curiously, the [Fe/H] distribution for NGC~2419 
derived from spectral synthesis of the Deimos data
is not as broad as the
[Ca/H] distribution.  The [Ca/H] measurements are slightly more uncertain
than the [Fe/H] measurements because the DEIMOS spectra contain fewer Ca
lines than Fe lines.  Figure~\ref{figure_kirby_hist} also includes the [Fe/H] and [Ca/H]
distributions for M13, a globular cluster known not to have a measurable
dispersion in elements heavier than Mg \citep[see, e.g.][]{sneden04,cohen05}.
The [Ca/H]
distribution in M13 is slightly broader than its [Fe/H] distribution, but
not as broad as the [Ca/H] distribution in NGC 2419.  These spectra for
these two clusters were obtained and the abundances were measured 
via spectral synthesis using the
same techniques.  Therefore, based on spectral synthesis we conclude
that the [Ca/H] distribution in NGC~2419 has a measurable dispersion 
whereas the dispersion in [Fe/H] is not obvious.  In this context one should
note the larger range and hence shallower slope (0.76 vs 1.23) in 
Fig.~\ref{figure_cah_hires_cat}
for [Ca/H] than for [Fe/H] found from the much smaller set of stars with
HIRES spectra.
This issue will be discussed
in more detail in a future publication.

\section{Discussion}

\begin{figure}
\epsscale{1.0}
\plotone{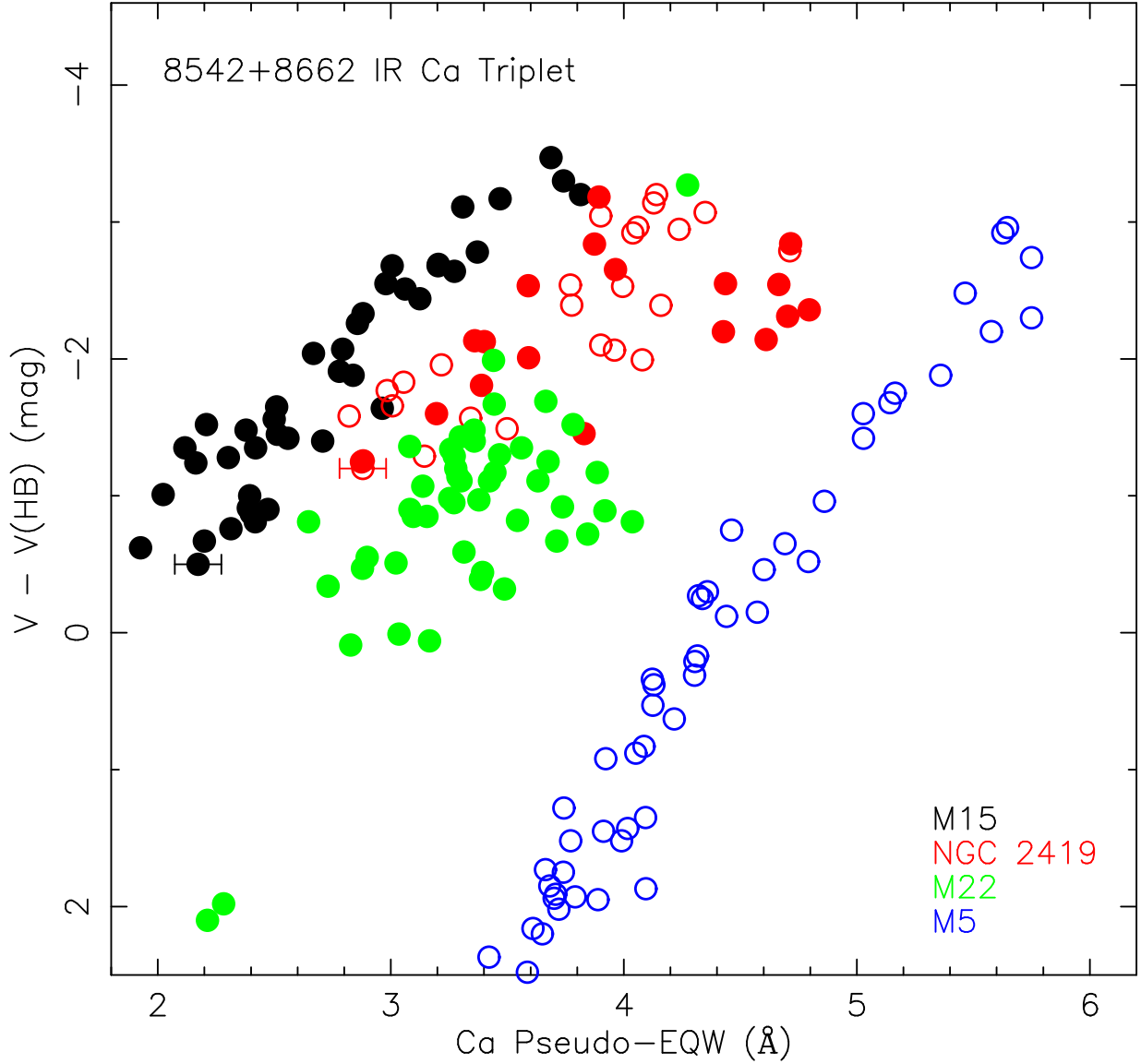}
\caption[]{The sum of the pseudo-equivalent widths for the two stronger lines
of the near-infrared Ca triplet are shown as a function of $M_V$
corrected for extinction and normalized with respect to the location
of the horizontal branch for four globular clusters
(M15, NGC~2419, M22, and M5, in order of increasing
metallicity), each shown
as a different color.  The NGC~2419 apparent members, shown in red,
are filled circles if the star was on both DEIMOS slitmasks, and
open circles if included in only one of the two slitmasks.  The
two stars whose membership in NGC~2419 is still uncertain are not shown.
Non-members have been carefully eliminated for NGC~2419 only.
As discussed in \S\ref{section_intro}, the
GC M22, which like NGC~2419, also shows a significant
spread in this figure, is known to have a small internal range in [Ca/H].
\label{figure_4gc_cat}}
\end{figure}

We have isolated a sample of 43 upper RGB stars in
the massive Galactic globular cluster NGC~2419, located far out
in the halo of the Milky Way at a distance of
84~kpc.  This sample
must consist almost entirely of members of NGC~2419,
as multiple lines of evidence demonstrate that
the expected number of halo field stars which could
meet our stringent requirements for membership
and lie within the area covered by our DEIMOS slitmasks
is less than one.  
Analysis of two DEIMOS 
slitmasks taken three years apart at the Keck~II telescope in 
the field of NGC~2419
demonstrates that these cluster members 
show a small but real range in the strength
of their near-IR Ca~II triplet lines
at a fixed $V$~mag
(see Fig.~\ref{figure_4gc_cat}).   We emphasize that we have
such data in hand for $\sim$10 Galactic GCs,
and that ignoring M22, NGC~2419 is the most
credible case for which star-to-star
variations of Ca abundances were seen.

\begin{figure}
\epsscale{1.0}
\plotone{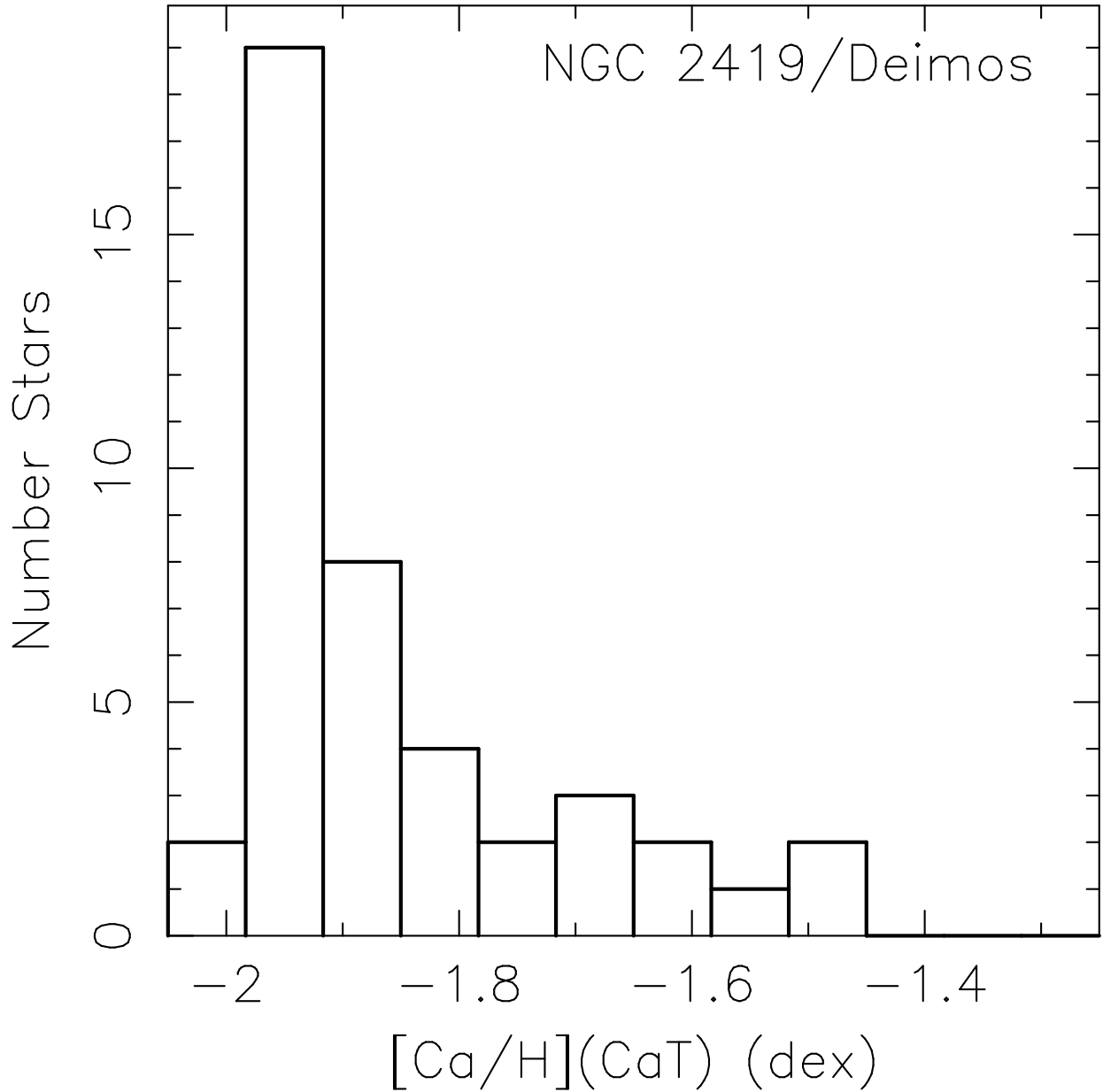}
\caption[]{A histogram of [Ca/H] inferred from their CaT
for the 43 members of NGC~2419.  The calibrating GCs are
M15, M53, and M13, with [Fe/H] assumed to be $-2.26$,
$-1.99$, and $-$1.54~dex respectively, adopted from the on-line
catalog of \cite{harris96}. A second order fit to our
Deimos spectra of samples of stars in each of the
three calibrators as a function of $V-V$(HB) is defined.  
Then for each of the NGC~2419 members, another second order fit 
to [Fe/H] as a function of CaT for the three calibrators at the luminosity 
$V-V$(HB) of the NGC~2419 star yields its [Fe/H].
[Ca/Fe] is assumed to be constant at +0.25~dex for all red giants in the
calibrating GCs.
\label{figure_feh_hist}}
\end{figure}

The range in NGC~2419 of CaT (the sum of the pseudo-equivalent
width of the two stronger lines of the near-infrared
Ca triplet) at a fixed $V$~mag
near the tip of the RGB is $\sim$1~\AA, 
from $\sim 3.8$ to $\sim 4.8$~\AA.  If we
assume the two Ca~II lines we use, at 8542 and 8662~\AA,
are on the damping part of the curve of growth,
then the Ca/H $\propto (\rm{CaT})^2$,
equivalent to a range of
0.2~dex in [Ca/H].  That is about the
range seen in [Ca/H] in detailed abundance analyses 
of high resolution spectra of 8 members of
this cluster that we have also completed.  The [Ca/H] histogram
for our sample of NGC~2419 giants 
deduced from our CaT measurements as calibrated by Deimos
slitmasks for M15, M53, and M5 (figure \ref{figure_feh_hist})
shows a strong peak in [Ca/H](CaT) at about $-1.9$~dex with a tail extending
to higher metallicities.
This result is supported by a spectral synthesis
analysis of the same Deimos slitmask following the techniques developed in
\cite{kirby08} and \cite{kirby10} which specifically excludes
the Ca triplet lines.
We therefore assume that the range in [Ca/H] in NGC~2419
is real.   A careful examination of the existing 
DEIMOS slitmasks in GC fields, and of those soon to be acquired,
for star-to-star variations in Ca abundance 
thus seems well justified, and is underway.

While many clusters show an intrinsic dispersion in lighter elements
up to and including Al \citep[see, e.g.][]{carretta09},
very few show a measurable dispersion in heavier
elements.  Dispersions in lighter elements could be explained by
self-pollution by AGB star winds, but spreads in Ca abundance must arise
from enrichment by SNe.  Based on the half-light radius in the Harris
catalog and the mass given by  \cite{sigmav09}, the gravitational
binding energy of NGC 2419 is about $4 \times 10^{51}$~erg.  A few core
collapse SNe would have been energetic enough to evacuate the cluster of gas
if NGC 2419 formed in isolation.  The system in which NGC~2419 formed must
have been massive enough to retain the SN ejecta.  We suggest that NGC~2419
formed in a now-disrupted dwarf galaxy.  The recent
theoretical model of GC formation by \cite{conroy10}
follows gas flows into and out of the young cluster in detail,
and can
reproduce many features of the chemical inventory of GCs
arising from multiple stellar generations, specifically the
correlated and anti-correlated variations among the light elements
and the relative fractions of these generations as observed today
\citep[see, e.g.][]{m15cn,carretta09}. 
However, even this model cannot
accomodate a spread in heavy element abundances without resorting
to a star cluster embedded in its own dark matter halo.

\cite{sigmav09} have ruled out more than $10^7~M_{\sun}$ of dark
matter within 500~pc of the center of NGC~2419.  Their upper limit depends
on the assumption of isotropic stellar orbits and no mass segregation.  In a
survey of 10 halo GCs, \cite{lane10}, extending via
their much larger samples earlier work, have found none with $M/L_V > 5$.
These measurements show that the central, most luminous regions of GCs,
including NGC~2419, are not dominated by dark matter.  However, it is
possible that the stellar component of NGC~2419 (and possibly other GCs)
formed in a dense dark matter halo but no longer reside at the center of the
halo or are significantly more compact than the dark matter distribution.

We ask how such a small, but non-zero range
in metallicity could have developed in such a metal-poor
system.  We require a situation where star formation
commenced, but before long, it was
quenched.  Since prompt initial enrichment is
characteristic of essentially all chemical
evolution models, this implies that almost all
the cluster gas was lost early, thus halting
star formation in the accreted satellite
dwarf galaxy whose central region eventually became NGC~2419.
How could this have happened given
that the GC is so far out in the halo at the present time ?
The answer may lie in its orbit, which \cite{pm2419}
suggest is extremely eccentric, with  pericenter at
$\sim$11~kpc and  apocenter at $\sim 90$~kpc. 
Such an orbit would be quite disruptive for the
gas, and 
star formation might cease completely after only
one or two orbital periods.  Since the period
calculated by \cite{pm2419} is short, only $1.2\pm0.5$~Gyr,
star formation in NGC~2419 could have been terminated quite rapidly.

\acknowledgements

We are grateful to the many people  
who have worked to make the Keck Telescope and its instruments  
a reality and to operate and maintain the Keck Observatory. 
The authors wish to extend special thanks to those of Hawaiian ancestry
on whose sacred mountain we are privileged to be guests. 
Without their generous hospitality, none of the observations presented
herein would have been possible.   We thank the anonymous referee
for helpful suggestions.
J.G.C. thanks NSF grant AST-0908139  for partial
support. Work by E.N.K. was supported by NASA through Hubble Fellowship
grand HST-HF-01233.01 awarded to ENK by the Space Telescope Science
institute, which is operated by the Association of Universities
for Research in Astronomy, Inc., for NASA, under contract NAS 5-26555. 
The analysis pipeline used to reduce the DEIMOS data was
developed at UC Berkeley with support from NSF grant AST-0071048.

{}

\clearpage

\begin{deluxetable}{l c rrr }
\tablenum{1}
\tablewidth{0pt}
\tablecaption{Data for Apparent Red Giant Members of NGC~2419
\label{table_sample} }
\tablehead{
\colhead{Name\tablenotemark{a}} & \colhead{V\tablenotemark{a}} & \colhead{CaT} &
\colhead{Detected Electrons} &  Masks\tablenotemark{c} \\
\colhead{} & \colhead{(mag)} & \colhead{(\AA)} &
\colhead{per pixel\tablenotemark{b}}    &   \colhead{} 
}
\startdata 
D6 &    18.457  &  4.08 & 6040 & 1 \\  
D17 &     18.681 &    2.98 & 6555 & 1\\
D18 &  17.921  &   4.00 & 10000 & 1 \\
S217 &    19.250 &   2.88 &   3080 &   1 \\   
S223 &    17.250  &  4.14 &   34050 &   1   \\ 
S243 &  19.202  &  2.87 &     12010 &   2  \\  
S357 &    18.996  &  3.83 &   13270 &  2  \\  
S406 &  17.797  &   3.96 &    61015 &  2  \\  
S457 &  18.317  &  3.36 &    34050 &  2  \\ 
S458 &  17.900  &  4.44 &    52378 &   2  \\  
S642 &  18.493  &  3.22 &    5945 &  1  \\  
S656 &  18.620  &  3.06 &    6195 &  1 \\ 
S704 &  18.308  &  4.61 &   33265   &   2  \\  
S731 &  18.793  &  3.01 &    4185 &  1  \\  
S759 &  18.883  &  3.34 &   3500 &  1  \\  
S809 &  17.915  &  3.59 &   51800 &  2  \\
S810 &  17.310  &  4.13 &   28200 &  1  \\  
S870 &  18.059  &  4.16 &    6310 &   1  \\ 
S890 &  18.960  &  3.50 &    3410 &   1 \\   
S934 &  18.324  &  3.40  &  29870 &   2  \\ 
S957 &  17.530  &  4.04 &   21400 &   1 \\  
S973 &  17.488  &  4.06 &   19380  &  1 \\ 
S1004 & 17.907  &  4.66 & 53380 & 2  \\
S1008 &   17.910  &  3.78 &  13660 &   1 \\ 
S1029 &   18.057  &  3.78 &  9430 &   1 \\
S1048 &  17.380  &  4.35 &   19940 &   1 \\
S1065 &  17.660 &   4.71 & 18080 &   1  \\
S1131 &  17.608  &   4.72 &   78300 &   2  \\
S1166 &   17.503 &  4.24 &  19130 &   1  \\ 
S1209 &  17.406  &   3.90 & 20730 &  1   \\  
S1215 & 18.850  &  3.20 &   18470 &   2 \\  
S1245 &  18.386  &  3.96 &  6180 &  1  \\
S1294 & 18.139 &   4.70 &    39580 &   2 \\   
S1305 &  17.611 &   3.87 &  74800 &   2 \\ 
S1306 & 18.091  &  4.80 &   36840 &  2   \\ 
S1340 &   19.161 &   3.14 &  2870 &   1  \\ 
S1391 &   18.441 &  3.59 &   28760 &  2  \\ 
S1497 &  18.643 &   3.39 &   22240 &   2  \\  
S1563 &    18.349  &  3.90 &  6225 &  1   \\ 
S1664 &  18.251 &   4.43 &   34800 &   2  \\  
S1810 & 19.194  &  2.88 &    2885 &   1  \\  
S1814 &  17.268  &  3.89 &    117100 &  2   \\ 
S1819 &   18.868  &  2.82 &    4000 &   1 \\  
S951\tablenotemark{d} &
        17.270  &  4.45 &   90860 &  2 \\  
S1673\tablenotemark{d} &        
        17.683  &  4.73 &  61010 &  2  \\  
\enddata
\tablenotetext{a}{Star IDs and $V$~mags are from the
database of \cite{stetson05}.}
\tablenotetext{b}{Total detected electrons/pixel for the two DEIMOS
slitmasks in the continuum near 8600~\AA.
At 8500~\AA, 1 pixel $\equiv 0.32$~\AA.}
\tablenotetext{c}{1 = star in one of two slitmasks, 2 = star in both.}
\tablenotetext{d}{These two stars may be members of 
of NGC~2419; see \S\ref{section_member}.}
\end{deluxetable}         

\clearpage

\end{document}